# Magnetofluidic mixing of a ferrofluid droplet under the influence of time-dependent external field

Sudip Shyam[1], Pranab Kumar Mondal[1]† and Balkrishna Mehta[2]

[1]Microfluidics and Microscale Transport Processes Laboratory, Department of Mechanical Engineering, Indian Institute of Technology Guwahati, Assam 781039, India

[2]Department of Mechanical Engineering, Indian Institute of Technology Bhilai, Raipur 492015, India

We report the experimental investigations on the mixing of a ferrofluid droplet with a non-magnetic fluid in the presence of a time-dependent magnetic field on an open surface microfluidic platform. The bright field visualization technique, in combination with the µPIV analysis, is carried out to explore the internal hydrodynamics of the ferrofluid droplet. Also, using the Laser-induced fluorescence (µLIF) technique, we quantify the mass transfer occurring between the two droplets, which in effect, determines the underlying mixing performance under the modulation of the frequency of the applied magnetic field. We show that the magnetic nanoparticles exhibit complex spatio-temporal movement inside the ferrofluid droplet domain under the influence of a time-dependent magnetic field, which, in turn, promotes the mixing efficiency in the convective mixing regime. Our analysis establishes that the movement of magnetic nanoparticles in presence of the time-periodic field strengthens the interfacial instability, which acts like a sparking agent to initiate an augmented mixing in the present scenario. By performing numerical simulations, we also review the onset of interfacial instability, mainly stemming from the susceptibility mismatch between the magnetic and non-magnetic fluids. Inferences of the present analysis, which focuses on the simple, wireless, robust, and low-cost open surface micromixing mechanism, will provide a potential solution for rapid droplets mixing without requiring pH level or ion concentration dependency of the fluids.

Email address for correspondence: mail2pranab@gmail.com; pranabm@iitg.ac.in

# 1. Introduction

With the advent of miniaturization, droplet-based microfluidics, which is ubiquitous in protein crystallization, biosensor, immunoassays, DNA-replication, cell-based assays, biomolecular extractions - to name a few - has emerged as an effective approach for precise manipulation of a discrete volume of fluid samples and analytes in recent years (Berry et al. 2011; Bogojevi et al. 2012; Mary et al. 2008; Shamsi et al. 2014; Tice et al. 2003; White et al. 2013; Zhang et al. 2011). One promising application of droplet-based microfluidics finds huge relevance to the micromixing technology, typically used in on-chip biochemical and biological analysis. The paradigm of droplet-based micromixing can be classified into two different types, namely open surface, and closed surface micromixing. Open surface droplet-based micromixing offers a few advantages over the closed surface microfluidic environment, such as simple fabrication, interfacial convenience, and free of any blockage such as bubble clogging (Greenspan 1978; Lin et al. 2017; Long et al. 2009; Pournaderi & Pishevar 2014; Smith 1995; Tam et al. 2009). It may be mentioned here that droplet-based mixing in the open surface microfluidic platform can be accomplished by using two approaches viz., passive approach, and active approach. The passive approach entirely sticks to the molecular diffusion between the phases to be mixed and can be tuned by altering the surface morphology (for example, patterned wettability controlled surface) towards achieving the desired controllability on the mixing (de Groot et al. 2016; Xing et al. 2011). While the active approach, which is more prevalent due to its reconfigurable flexibility, primarily relies on external force fields such as electric field, magnetic field, acoustic waves, and light energy for maneuvering the droplet flow field (Behera et al. 2019; Grassia 2019; Meng & Colonius 2018; Shang et al. 2017; Singh et al. 2018).

Of all these active approaches, utilization of the magnetic field has evolved as a promising technology in the paradigm of droplet-based mixing in microfluidic platforms, attributed primarily to its inherent advantageous features. A few of such notable features include biocompatibility, ease of incorporation, low cost, less invasive, no induction heating, precise manipulation of the contact line, and many more (Huang et al. 2017; Liu et al. 2018; Nguyen 2012; Zhang & Nguyen 2017). It is worth mentioning here that researchers have exploited the flexibility of magnetic manipulation in the open surface microfluidic framework for achieving controlled mixing (Biswal & Gast 2004; Franke et al. 2009; Lee et al. 2009; Martin et al. 2009; Roy et al. 2009; Sing et al. 2010; Vilfan et al. 2010). Ferrofluid is a colloidal suspension of nanoparticles in a non-magnetic carrier medium (Odenbach 2002; Rosensweig 1984). The nanoparticles are usually stabilized by surfactants such that they exhibit continuum behavior in the presence of a strong magnetic field. Due to the superparamagnetic nature of the nanoparticles, this typical fluid has attracted significant attention in the scientific community because of its promising potential in the area of magnetofluidic based applications (Afkhami et al. 2008, 2010; Ganguly et al. 2004; Mahendran & Philip 2012; Qiu et al. 2018; Rowghanian et al. 2016; Shyam et al. 2019, 2020b, 2020a; Vieu & Walter 2018). The rapid response of the nanoparticles to the magnetic field offers tremendous flexibility in stirring/mixing related applications in lab-on-a-chip (LOC) device/systems (Hejazian et al. 2016; Nouri et al. 2017; Tsai et al. 2009; Wang et al. 2008; Zhu & Nguyen 2012). Applications of the magnetic field ensure the development of instability at the liquid-liquid interface due to the magnetic susceptibility mismatch, and the consequence of this phenomenon results in an enhanced mixing (Zhu & Nguyen 2012). The interfacial instability gets further aggravated under the influence of the time-periodic magnetic field owing to the various involved spatio-temporal scales. It may be mentioned in this context here that a few researchers have explored the implications of the time-periodic magnetic field on the mixing characteristics between two fluids as well (Wang et al. 2008; Wen et al. 2009). The time-periodic magnetic field perturbs the flow domain by ensuring a transient interactive force through magnetophoresis, which, in turn, enhances the mixing between the fluids appreciably.

Albeit several underlying issues of the magnetic field modulated micromixing analysis have been well explored, see Refs. (Gao et al. 2015; Hejazian et al. 2016; Munaz et al. 2017; Nouri et al. 2017; Roy et al. 2009; Tsai et al. 2009; Wang et al. 2008; Wen et al. 2009; Zhu & Nguyen 2012), the phenomenon of magnetophoresis, which is effectively used in manipulating microflows, on the droplet-based micromixing is sparsely studied. It may be mentioned in this context here that in the paradigm of droplet-based mixing, the rotating magnetic field has been used in a synergetic way for augmenting the mixing appreciably (Gao et al. 2015; Munaz et al. 2017; Roy et al. 2009). The rotation of the magnetic field leads to the generation of rotating magnetic chains, which further promotes the mixing phenomena inside the droplet domain following the magnetoconvection effect (Roy et al. 2009). Although a rotating magnetic field ensures a significant augmentation of mixing in the droplet domain, its application in practice complicates the design as well as the fabrication process of the open surface droplet-based micromixer. This aspect, to be precise, limits the applicability of the rotating magnetic field in the area of open surface micromixing platform. The droplet-based micromixing under the influence of the time-dependent magnetic actuation could be an interesting proposition, attributed to the rich physical interplay of various spatio-temporal scales involved, as well as to its immense consequences on the efficient mixing following interfacial instabilities. This aspect has not been studied in the literature to date.

In this work, we experimentally explore a new method of generating strong convection inside the ferrofluid droplet under the modulation of a time-periodic magnetic field. We place the ferrofluid droplet (base droplet) in between two alternatively acting electromagnets. We show that the transiences in nanoparticles induce a magnetoconvective flow, which, in turn, promotes the mixing of the base droplet with the non-magnetic sister droplet being injected from the top. We show that the intermittent motion of the magnetic nanoparticles under the influence of the time-dependent magnetic actuation triggers the interfacial instabilities to a significant extent. This phenomenon eventually brings about sufficient agitation in the bulk liquid of the droplet, leading to enhanced mixing. Also, we numerically simulate the concentration field in the droplet domain under the influence of the magnetic field for the qualitative understanding of the mixing characteristics. In what follows, we divide this study into three sections. In the first section, we explore the internal convective characteristics of the ferrofluid droplet in the presence of a magnetic field with the help of bright field visualization. In the intermediate section, following the µPIV measurement technique, we explore the internal hydrodynamics of the bulk liquid inside the droplet in the presence of a magnetic field. In the final section of this article, we explore the implication of this augmented advective force on the ferrofluid droplet mixing with another non-magnetic droplet.

## 2. Materials and Methods

### 2.1. *Fluid characterization and substrate preparation*

The preparation of the ferrofluid solution is an involved procedure. The first step of this procedure is the synthesis of the Iron oxide ($Fe_3O_4$) nanoparticles. For the present analysis, we stick to the co-precipitation method for the chemical synthesis of $Fe_3O_4$ from an aqueous mixture of $Fe^{+3}/Fe^{+2}$ (2:1). For the sake of conciseness in the presentation, we do not discuss here the remaining steps. The interested readers may refer to the recent work from our group for the detailed discussion of the involved stages of this process (Shyam et al., 2020a, 2020b). Figure 1(a) shows the magnetization curve of the prepared ferrofluid solution. The magnetization curve of the ferrofluid solution was generated with the help of a vibrating sample magnetometer (VSM). The saturation magnetization of the ferrofluid solution was found to be around 0.12 emu/gm. The inset at the left-top corner of figure 1(a) shows the distribution of the size of the nanoparticles. In contrast, the inset at the right-bottom side shows the absolute zeta potential of the ferrofluid solution, as was measured by DelsaNano-C. The average size of the nanoparticles was found to be around $50 \pm 2$ nm, whereas the measured value of absolute zeta potential of the ferrofluid solution was approximately 53 mV. Note that this typical value of zeta potential is suggestive of an electrostatically stable solution (Xu 2002).

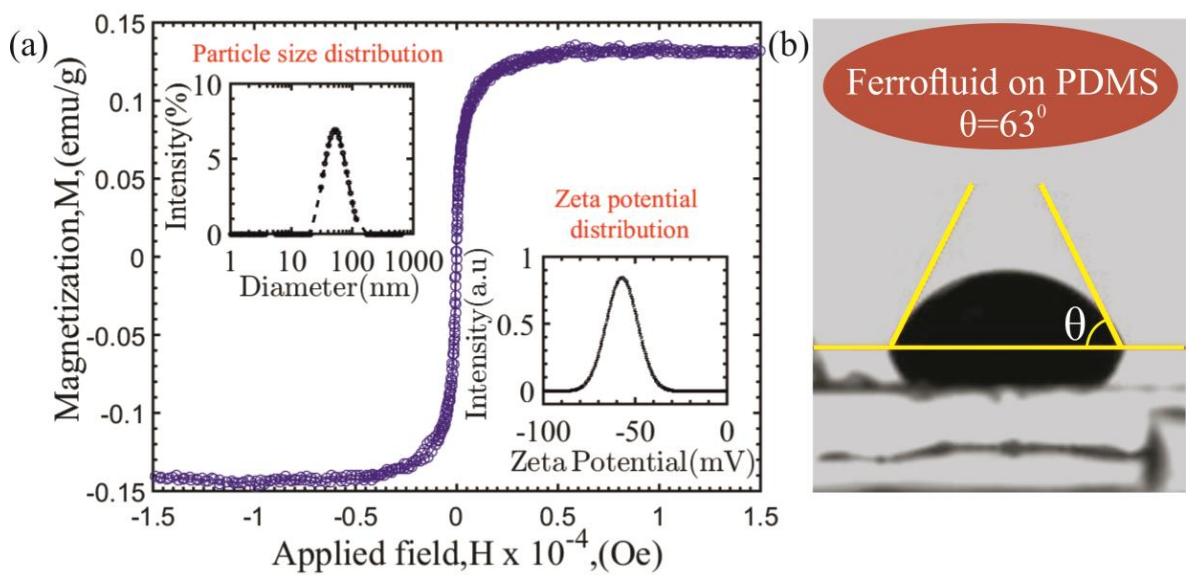

**FIGURE** 1. (**Color online**) (a) Plot depicts the magnetization curve of the prepared ferrofluid sample, as was measured by VSM. The left-hand side inset illustrates the variation of the magnetic nanoparticle size distribution. The right-hand side inset shows the variation of the zeta potential of the ferrofluid solution. The volume fraction of the magnetic nanoparticles (MNPs) in the base fluid was around 0.1%. (b) Plot illustrates the static contact angle of a sessile ferrofluid droplet on the treated PDMS substrate.

Microscopic glass slides (Make: Struers) of 1.1 mm thickness and 27 x 46 mm$^2$ in size were used as a substrate. The glass slides were coated with a thin cured PDMS layer to prepare the final hydrophobic substrate. The PDMS solution was prepared by mixing silicone elastomer (Make: SYLGARD 184) with a curing agent in the ratio of 10:1. The solution was de-gasified in a vacuum chamber. The degassed solution was then poured on to the glass substrate and coated by a spin coater (Make: Apex instruments) at 3400 RPM for the 50 s. The spinning effect is resulting in a thin as well as uniform deposition of PDMS layer on the substrate. The prepared glass substrate was then placed on a hot air oven for around 3 hours at a constant temperature of 80 °C. The ferrofluid solution forms a contact angle of $\theta_{ferrofluid} \sim 63°$ on the PDMS substrate. The magnetic nanoparticles are coated with surfactant (Lauric acid) essentially to prevent any interparticle agglomeration in the ferrofluid solution. Note that the coated surfactant layer lowers the contact angle of the ferrofluid solution to the aforementioned value, as witnessed in figure 1(b) mentioned above.

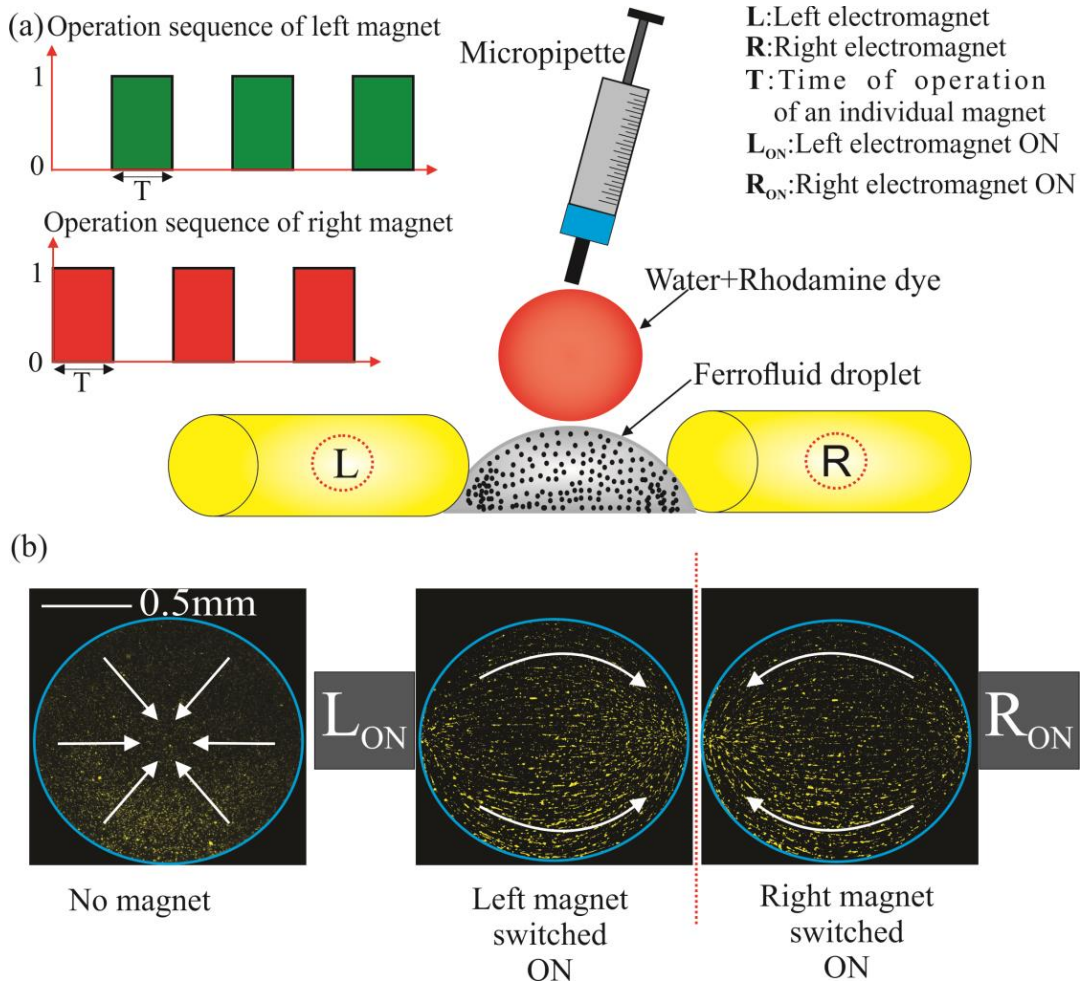

**FIGURE 2. (Color online)** (a) Schematic representation of the working mechanism of the proposed microfluidic platform for rapid and efficient droplet mixing. A fluorescent water droplet is injected from top to a sessile ferrofluid droplet under the actuation of a time-dependent magnetic field. The sequence of operation of the electromagnet is shown in the top left corner of the figure. When the left magnet is in ON-state, the right magnet remains in OFF-state and vice versa. All the involved symbols were defined aptly on the top right-hand side of the figure. (b) Plots show the motion of the fluorescent seeded particles in the presence and absence of the magnetic field. The white-colored arrow shows the direction of the bulk fluid flow inside the droplet.

## 2.2. Magnetic forcing actuation setup

For the application of magnetic fields, we fabricate two electromagnets by winding 26 SWG enamel coated copper wires (~ 40 turns per cm) over 6 mm diameter and 100 mm long iron cores. We place the electromagnets at a distance of 0.2 mm from the periphery of the droplet. An electric current of required strength from a DC power source (Make: Aplab) is supplied to the electromagnets for its activation. We integrate an in-house developed circuit in the power supply line to provide the pulse current to the electromagnets. The pulsed current establishes the time-dependent operation (right: "ON"/left: "OFF" and vice-versa) of the fabricated electromagnets. Note that the time-dependent operation ensures alternate actuation of the electromagnets at a predefined specific instant of time. In the present study, we keep the magnetic field flux density constant at $\bar{B} = 400\ G$, while the actuation frequency varies from $f = 0.3\ Hz$ to $5\ Hz$. In the supplementary section of this paper, we have provided the distribution of this magnetic field generated by the electromagnet inside the droplet domain (refer section 1 of the supplementary material). To obtain the spatial variation of the applied forcing, we numerically simulate the magnetic field distribution of fabricated electromagnet using COMSOL Multiphysics®.

## 2.3. Experimental setup and the working principle

We show, in figure 2(a), the complete methodology of the experiments conducted in this study through a series of schematic depictions. We now briefly discuss the experimental procedure for the sake of completeness and ease in the understanding of the readers. A ferrofluid droplet of volume 1 µl is placed on the treated PDMS substrate using a digital microdroplet dispenser (Make: Tarsons). We perturb the ferrofluid droplet by a time-dependent magnetic actuation, originating from two axially aligned electromagnets (cf. figure 2(a)). The controlled magnetic perturbations generate internal convections inside the droplet. The imposed magnetic actuation strength for about 40 s leads to the formation of a chain-like structure of nanoparticles inside the ferrofluid droplet. Following this event of chain formation, a water droplet containing fluorescent dye (0.05g of Rhodamine 6G in 20 ml of De-Ionized water) of the equal volume is injected on to the ferrofluid droplet with the help of a microdroplet dispenser. We use a hall probe digital gaussmeter (Make SES instruments) to measure the strength of the applied magnetic field. Note that we perform all experiments for base magnetic field strength, $\bar{B} = 400\ G$. Also, care has been taken during experiments to isolate the droplet from the convection currents of the surrounding air such that no shear-induced mixing takes place.

Figure 3(a) shows the schematic of the experimental setup. As already discussed, the present experimental study is divided into three primary parts: the bright field visualization, the µPIV (micro-particle image velocimetry) analysis, and the µLIF (micro laser-induced fluorescence) investigation. Also, we numerically simulate the mixing process in the present problem, primarily to compare the insights gained from the experimental observations. It is worth mentioning here that the intricate details captured from the simulations cooperate in explaining the experimental results for a better understanding of the flow physics of our interest. In the bright-field visualization, white light from the mercury lamp illuminates the droplet flow field. We observe the transmitted white light from the bottom of the substrate with the help of a 10X (magnification) objective lens having a numerical aperture (NA) of 0.24. In the bright-field visualization, we perceive the motion of the magnetic nanoparticles (MNPs) in the presence of a time-dependent magnetic field. We use an objective lens of higher magnification of 20X to capture the chain-like cluster[1] of the magnetic nanoparticles.

We perform µPIV investigation for the quantification of the internal flow hydrodynamics of the droplet. The µPIV experimental setup consists of three main components: (a) an inverted microscope (Leica: DM IL LED), (b) a monochromatic light source, and (c) a camera. The ferrofluid droplet is seeded with 1µm diameter fluorescent microspheres (Make: Molecular Probes Inc.). The fluorescent particles ensure an acceptable low noise level under the present illumination conditions. In figure 2(b), we show the distribution of the fluorescent particles in the droplet domain both in the presence and absence of a magnetic field. We keep the droplet–electromagnet assembly over the stage of the inverted microscope, and the droplet is illuminated by monochromatic light. An epifluorescent prism filter is used on the optical path to eliminate the background light. Double images are captured per realization in such a way that the seeding particles move approximately 1/4$^{th}$ size of the interrogation window. For the calculation of the instantaneous velocity vector field of the present setup, we use a cross-correlation algorithm with an interrogation window of size 64 × 64 pixels[2,] and a 50% overlapping between each window.

---

[1] The formation of chain-like cluster is an important event associated with the internal hydrodynamics of the ferrofluid drop. This feature has been elaborated in the results and discussion part of this article in greater detail.

We use PIVLab to analyze the captured images for the assessment of the velocity field (Thielicke & Stamhuis 2014). Before using the cross-correlation algorithm, we take the raw µ-PIV images in ImageJ software to obtain the overlapped images (Schneider et al. 2012). Note that the process of overlapping ensures an increased number of seeding particles per interrogation window, resulting in an easy peak detection (during the cross-correlation algorithm).

We appeal to the µLIF investigation for the quantification of the underlying mixing phenomena between the base (ferrofluid) and sister (water) droplets. The experimental setup is similar to the µPIV configuration. For the present task (µLIF investigation), the ferrofluid droplet domain is illuminated by the fluorescent light, and the water droplet (containing fluorescent dye) is subsequently injected from the top with the help of a microdroplet dispenser (Make: Tarsons). Since the beginning of the water droplet injection, we start recording the distribution of the water droplet inside the ferrofluid droplet domain. The recorded images are subsequently converted to the grayscale format. For the elimination of noise from the captured images, we subtract the intensity histogram of the base reference image from all the subsequent recorded images for a particular experiment. Following this, we calculate the standard deviation of the pixel intensity in the droplet region, as given by,

$$C' = 1 - \sqrt{(1/N) \sum_{1}^{N} [(P - \bar{P})^2 / \bar{P}^2]} \qquad (2.1)$$

Where P, is the intensity which varies from 0 to 256, while the mean pixel value ($\bar{P}$) is given by:

$$\bar{P} = (1/N) \sum_{1}^{N} P \qquad (2.2)$$

$N$ represents the number of pixels. Initially, when the droplet is in the unmixed state, $C' = C_0$ and at the final stage of mixing, $C' = C_\infty$. Thus, the mixing index is normalized as

$$\bar{C} = \frac{C' - C_0}{C_0 - C_\infty}; 1(= \text{Mixed state}) < \bar{C} > 0(= \text{Unmixed state}) \qquad (2.3)$$

During experiments, we ensure to maintain the temperature and the humidity inside the laboratory at $25 \pm 0.5°C$ and $67 \pm 1\%$, respectively. The calculated value of the Bond number ($Bo$) is found to be less than one. However, the magnetic bond number is calculated as $Bo_m = \mu_0 H^2 R/\gamma$ where $\mu_0, H, R,$ and $\gamma$ represents the magnetic permeability of vacuum, magnetic field intensity, radius of the droplet, and the interfacial tension, respectively. For the present case $Bo_m > 1$, which signifies the dominance of the magnetic force on the droplet domain. For ensuring repeatability, we perform each experiment four times using the same sample. It is worth mentioning here that the maximum uncertainties involved in each run do not exceed 8%.

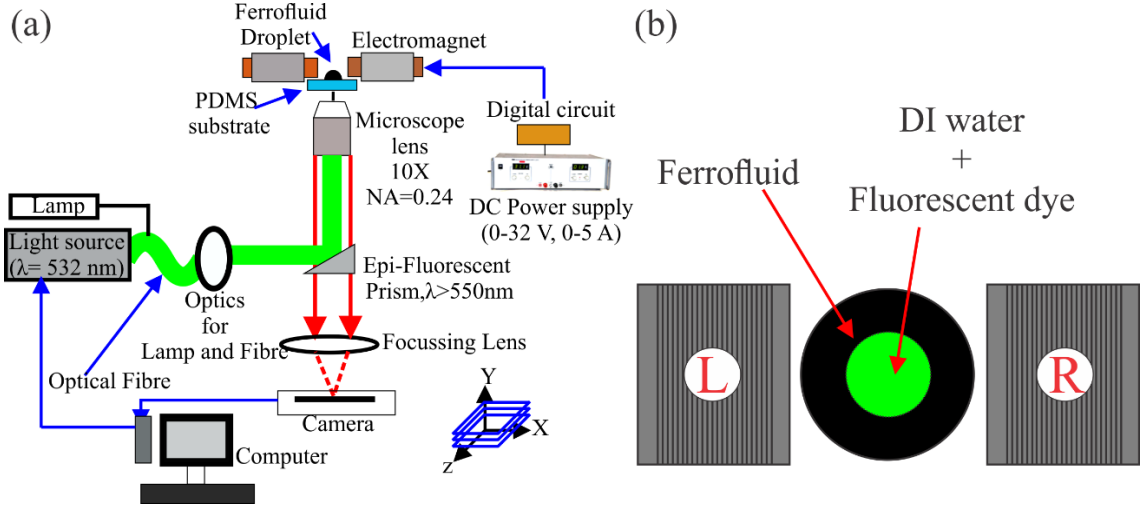

**FIGURE** 3. **(Color online)** (a) Schematic of the experimental set up along with its components. The experimental setup is used to conduct the bright field visualization, micro-particle image velocimetry (μPIV) analysis, and micro-laser induced fluorescence (μLIF) investigation in the droplet domain. All the components are aptly described in the text. (b) Schematic of the simulated two-dimensional computational domain. L and R indicate the left and right magnet, respectively. (Schematic is drawn not to scale)

### 2.4. *Numerical method*

We also make an effort to simulate the droplet flow field numerically, essentially to understand the intricate dynamics of the convective-diffusive mass transfer between the two droplets. We perform in this study two-dimensional (2D) numerical simulations using COMSOL Multiphysics®. We refer to figure 3(b), showing the schematic of the computational domain consisting of the sessile droplet (the base droplet), the sister droplet, and the two time-dependent magnets. The continuity and the Navier-Stokes equations for the unsteady, viscous and incompressible fluid flows are given by:

$$\frac{\partial \rho}{\partial t} + \nabla \cdot (\rho \bar{u}) = 0 \qquad (2.4)$$

$$\frac{\partial(\rho\bar{u})}{\partial t} + (\bar{u}.\nabla)\bar{u} = -\nabla P + \nabla\{\eta(\nabla\bar{u} + \nabla\bar{u}^T)\} + \bar{F} \tag{2.5}$$

Where $\rho$ is the density($Kg/m^3$) of the fluid, $\bar{u}$ is the velocity field($m/s$), $\eta$ is the viscosity of the fluid ($Pa - s$), and $\bar{F}$ is the volume force in ($N/m^3$). We calculate the magnetic field acting on the droplet flow domain by solving the Maxwell equations as given by (Griffiths & Inglefield 2005):

$$\nabla \cdot \bar{B} = 0 \tag{2.6}$$

$$\nabla \times \bar{H} = 0 \tag{2.7}$$

Where $\bar{B}$ is the magnetic flux density and $\bar{H}$ is the intensity of the magnetic field. The magnetic flux density ($\bar{B}$) is given by (Griffiths & Inglefield 2005):

$$\bar{B} = \mu_0(\bar{H} + \bar{M}) \tag{2.8}$$

$\mu_0$ is the permeability of vacuum, $\bar{M}$ is the magnetization of the flow domain. The magnetization ($\bar{M}$) is described as (Griffiths & Inglefield 2005):

$$\bar{M} = \chi\bar{H} \tag{2.9}$$

Where $\chi$ is the magnetic susceptibility of the fluid. Note that $\bar{F}$ in (2.5) is the force that comprises of the gravity force $(\bar{F}_g)$, interfacial tension $(\bar{F}_s)$, and the magnetic force $(\bar{F}_m)$. The gravity force is neglected since the Bond number is less than one ($< 1$). The interfacial tension can also be ignored since the fluids are miscible. The magnetic force ($\bar{F}_m$) is calculated from (Strek 2008):

$$\bar{F}_m = (\bar{M} \cdot \nabla)\bar{B} \tag{2.10}$$

Using $\bar{B} = \nabla \times \bar{A}$ in (2.10), the component of the magnetic force ($\bar{F}_m$) in $X$ and $Y$ direction respectively, can be written as (Nouri et al. 2017),

$$F_{m,x} = \frac{C\chi}{\mu_0\mu_r^2}\left[\frac{\partial A_z}{\partial y} \cdot \frac{\partial^2 A_z}{\partial x \partial y} + \frac{\partial A_z}{\partial x} \cdot \frac{\partial^2 A_z}{\partial x^2}\right] \tag{2.11a}$$

$$F_{m,y} = \frac{C\chi}{\mu_0\mu_r^2}\left[\frac{\partial A_z}{\partial x} \cdot \frac{\partial^2 A_z}{\partial x \partial y} + \frac{\partial A_z}{\partial y} \cdot \frac{\partial^2 A_z}{\partial y^2}\right] \tag{2.11b}$$

Where $\bar{A}$ is the magnetic vector potential $(Wb/m)$, $\mu_r$ is the relative permeability of the magnet. The advection-diffusion equation governing the mass transfer phenomena in the droplet domain (modeling framework) is given by,

$$\frac{\partial C}{\partial t} + (\bar{u}.\nabla C) = \nabla \cdot \{D\nabla C\} \tag{2.12}$$

Where $C$ is the concentration of the fluid $(mole\ /\ m^3)$, and $D$ is the diffusion coefficient of the fluid $(m^2\ /\ s)$. The advective-diffusion process will alter the density and viscosity of the mixture in a time-dependent magneto convective flow. Therefore, the density and viscosity of the fluid need to be represented as a function of the concentration flow field (Wen et al. 2011),

$$\rho_{mix} = C\rho_f + (1-C)\rho_w \tag{2.13}$$

$$\eta_{mix} = \eta_f e^{R(1-C)} \tag{2.14}$$

$$\text{Where, } R = ln(\eta_w/\eta_f) \tag{2.15}$$

Where the subscripts $mix$, $f$, and $w$ stand for mixture, ferrofluid, and water, respectively. Similarly the effective density $(\rho_f)$ and viscosity $(\eta_f)$ of ferrofluid can be calculated as (Brinkman 1952),

$$\rho_f = \psi\rho_{MNP} + (1-\psi)\rho_w \tag{2.16}$$

$$\eta_f = \eta_w \left(\frac{1}{(1-\psi)^{0.25}}\right) \tag{2.17}$$

$\psi$ is the volume fraction of magnetic nanoparticles (MNPs) in the ferrofluid solution. In the numerical simulations, the boundary and the initial conditions are as follows: no-slip and no flux boundary conditions at the outer boundary of the ferrofluid droplet; continuity in flux at the interface of the two droplets. For the magnetic field simulations, we consider the computational domain to be large enough, enabling us to apply the magnetic insulation boundary condition at the boundaries.

## 3. Results and discussion

The ferrofluid droplet flow domain consists of a magnetic part and a non-magnetic part. The magnetic nanoparticles (MNPs) constitute the magnetic part, while the non-magnetic part comprises of the bulk carrier liquid. As already mentioned, we explore the motion of the MNPs by bright-field investigations, while the μ-PIV measurement technique quantifies the bulk flow motion. The movement of the nanoparticles under the influence of the applied magnetic field alters the flow dynamics inside the ferrofluid droplet domain, which in turn, changes the concentration field and results in better mixing. We also numerically simulate the magneto convective flow and its effect on the concentration field. We will next discuss systematically the underlying issues of the flow dynamics in the presence of a magnetic field and subsequent mixing in the forthcoming sections.

### 3.1. *Droplet Internal hydrodynamics*

*3.1.1. Bright field visualization*

In figures 4(a)-(b), we show the movement of the MNPs when a time-dependent magnetic field of frequency $f = 0.3$ Hz perturb the ferrofluid droplet domain. A magnetic field frequency of 0.3 Hz implies that the time period of the particular magnetic forcing cycle is of 3.333 s. Note that out of the total cycle time (~3.333 s); the right magnet remains in ON state for 1.6667 s, while the left magnet is in ON state for another 1.6667 s. This process is repeated in the time-periodic forcing environment. Since both the electromagnets are aligned along the diametrical direction of the droplet, we show, in figures 4(a)-(b), the snapshots for the cases when the right magnet is active. Needless to say, the MNPs motion will exhibit qualitative similar kinetics even when the left magnet is turned on to the active state, albeit the direction of the MNPs motion would change. Figure 4(a) demonstrates the bright field visualization of the motion of the MNPs at various temporal instants. Note that zero '0' ms denotes the state when the right magnet is switched ON. The MNPs on the realization of the applied magnetic force start migrating towards the active magnet (Right magnet). Precise observation of figure 4(a) shows that the MNPs moves in the droplet flow field following the formation of the cluster having a head and a long tail. The head of the cluster moves towards the active magnet (right magnet) upon piercing through the carrier liquid. This typical piercing action of the cluster (of MNPs), in turn, creates agitations inside the bulk liquid of the ferrofluid droplet. The course of this agitation endorses a motion in the droplet domain. Note that the induced motion due this agitation is in the opposite direction of the moving MNPs motion (as shown by green arrows in figure 4(a)). The cluster (of MNPs) on reaching the vicinity of the active magnets strikes the triple contact line. Following which it undergoes deformation and in-process rearranges itself according to the prevailing magnetic force environment. Subsequently, the MNPs agglomerates and realigns in a chain-like formation. It is worth mentioning here that this chain-like cluster formation is primarily due to the interparticle dipole-dipole interaction existing between the MNPs in the presence of the magnetic field, as given by $I_m(ij) = -\left[3\frac{(m_i \cdot r_{ij})(m_j \cdot r_{ij})}{r_{ij}^5} - \frac{m_i m_j}{r_{ij}^3}\right]$, where $r_{ij} = r_i - r_j$ is the distance between the $i_{th}$ and $j_{th}$ nanoparticles having magnetic moments $m_i$ and $m_j$ respectively (Mendelev & Ivanov 2004). This chain-like cluster of the MNPs breaks down as the inactive magnet (the left magnet for the present configuration) returns to its active phase, and the motion of the deagglomerated MNPs continues in a similar manner towards the active magnet (Left at the prevailing situation). This typical fashion of movement of the MNPs inside the droplet domain, i.e., with head and tail, is encountered only at low frequency. At higher frequencies, such kind of motion of the MNPs is not observed. We will discuss this non-intuitive behavior in the latter part of this section in greater detail.

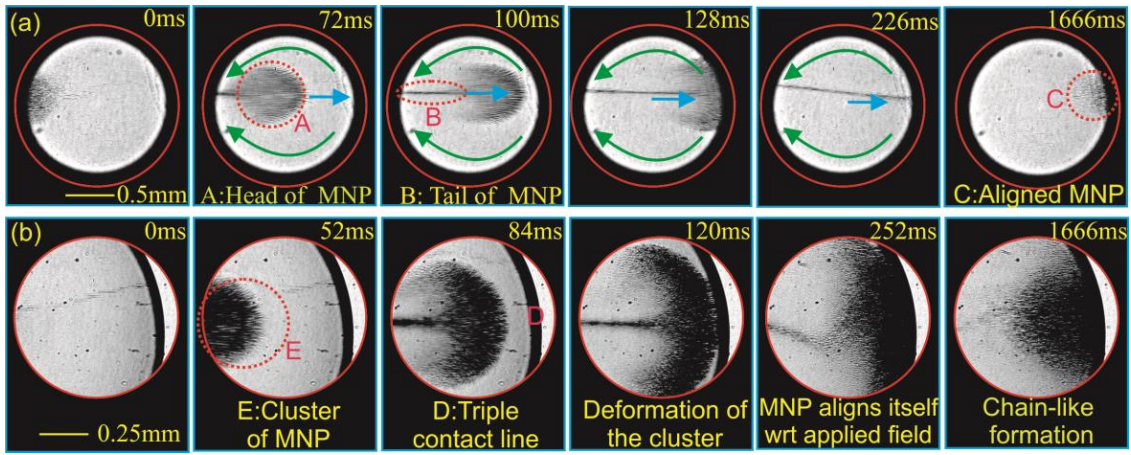

**FIGURE** 4. **(Color online)** (a) Snapshots depict the motion of the magnetic nanoparticles (MNPs) inside the ferrofluid droplet domain at various temporal instances when perturbed by the time-dependent magnetic field frequency ($f$) of 0.3 Hz. The blue-colored arrows indicate the direction of the MNPs. The green-colored arrows show the direction of the motion of the bulk carrier fluid. The images are recorded at a microscope magnification of 10X. (b) Snapshots illustrate the spatio-temporal motion of the migrating MNPs near the magnetically active triple contact line. The images are recorded at a higher microscope magnification of 20X.

To further understand the arrangement of the magnetic nanoparticles near the triple contact line area of the active magnet, we make an effort in figure 4(b) to demonstrate the flow field captured with a higher magnification of 20X. A closer observation of figure 4(b) shows that the head of the clustered MNPs undergoes deformation on striking the active triple contact line (at around 120ms). As a result of this deformation, we observe the re-alignment of the MNPs in tune with the applied magnetic force field, leading to the development of a chain-like cluster. Note that the individual MNP acts as a dipole in the presence of the applied magnetic field, which, in effect, leads to the development of the chain-like cluster formation.

In figure 5, we show the spatio-temporal variation of the MNPs inside the droplet domain when perturbed by a time-dependent magnetic field of four different frequencies $f = 0.3\ Hz$, $1\ Hz$, $3\ Hz$, $and$ $5\ Hz$ s respectively. On actuation of the electromagnet, the MNPs migrate towards the active electromagnet following the typical cluster-like formation, as already explained in the preceding discussions, i.e., a head moves in the forward direction followed by a long tail. However, this distinctive motion (of MNPs) in a clustered fashion is limited to the lower frequencies, i.e., for $f = 0.3\ Hz$ and $1\ Hz$. At a higher frequency of the magnetic field $f = 3\ Hz$ and $5\ Hz$, the presence of neither the head nor the tail (of the MNP's) could be traced in the domain, as can be seen from figure 5. In addition to that, a distinct chain-like cluster formation is also not observed at a relatively higher frequency (precisely $f = 5\ Hz$ case). We would like to discuss another interesting observation on the non-dimensional time $(t^*)$ taken by the MNPs cluster to reach the triple contact line nearer the active electromagnet as follows. The non-dimensional time $(t^*)$ is defined as the ratio of instantaneous time to the time period for which an individual magnet remains at ON state $(T)$, specifically, $t^* = t/T$. At low frequency, the MNPs could reciprocate between the two magnetically active zones, as can be seen from figure 5. However, at higher frequencies, the MNPs are unable to reach the magnetically active zones, as observed in figure 5. In order to figure out the underlying physical reasoning behind this observation, we look at the effects of the advective time scale of the MNPs and the magnetic perturbation time scale of the electromagnet. The advective time scale $(t_u = D_h\ /\ U_{MNP})$ refers to the time taken by the MNPs to travel the characteristics length, i.e., the droplet diameter, at a particular strength of the actuation force. While the perturbation time scale $(t_m = 1/2f)$ implies the time over which an individual magnet remains in the ON stage. For the calculation of the advective time scale, we tracked the MNPs motion with ImageJ plugin Trackmate® (Schneider et al. 2012). Following which the average velocity $(U_{MNP})$ of the magnetic nanoparticles was found to be around 11mm/s. Based on this average velocity, the advective time scale $(t_u = D_h\ /\ U_{MNP})$ of the MNPs is found to be around 0.13 s. On the other hand the magnetic perturbation time scale, $t_m = 1/2f$ becomes 1.667 s, 0.5 s, 0.1667 s, and 0.1 s for 0.3 Hz, 1 Hz, 3 Hz, and 5 Hz cases, respectively. It is because of this imbalance between the advective and magnetic perturbation time scale ($t_u < t_m$), at lower frequencies of the magnetic field, particularly for 0.3 Hz and 1 Hz, the MNPs could reciprocate between the two magnetically active zones, as can be seen from figure 5, whereas for a relatively higher frequency, i.e., for $f = 5\ Hz$ case, the advective time scale of the MNPs is higher as compared to the magnetic perturbation time scale. As a consequence, the MNPs could not reciprocate between the two magnetically active zones. However, at the magnetic field frequency of 3 Hz, the advective time scale is almost balanced by the perturbation time scale ($t_u \sim t_m$). Quite notably, at this frequency, the MNPs travel in the most optimal way (see movie 1 given in the supplementary information section). This particular optimal motion of the MNPs at the magnetic field frequency of 3 Hz has a huge implication in the internal convections of the bulk carrier fluid flow, as well as its significance on subsequent mixing. We will discuss this part in the forthcoming sections.

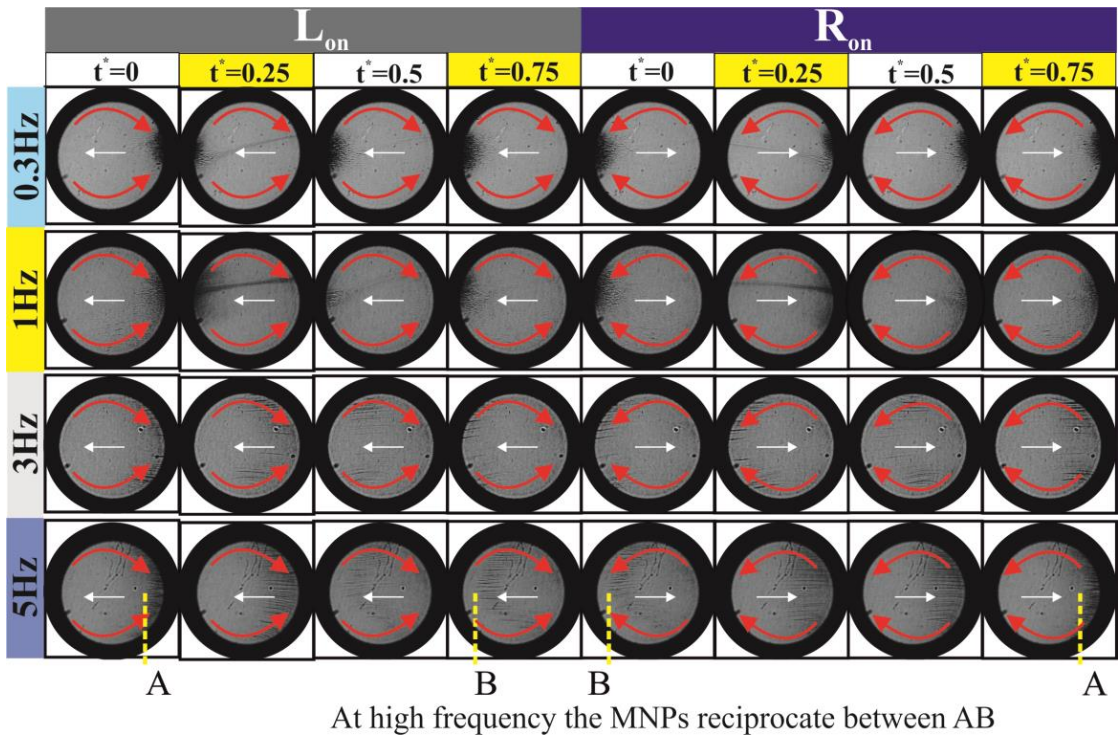

**FIGURE 5. (Color online)** Plot depicts the motion of the magnetic nanoparticles (MNPs) under the influence of magnetic field frequency of 0.3 Hz, 1 Hz, 3 Hz, and 5 Hz, respectively, for the various time instances of functioning of the electromagnet. The white colored arrows indicate the direction of the MNPs motion, and the red colored arrows show the direction of the bulk flow. "AB" denotes the position between which the MNPs reciprocate at $f = 5\ Hz$. $t^* = t/T$, where $t$ is the instantaneous time and $T$ is the time of operation of an individual magnet.

*3.1.2. µ-PIV investigation*

We have seen in the previous section that the advective time scale ($t_u$) of the MNPs gets almost balanced by the perturbation time scale ($t_m$) of the electromagnet when the magnetic field frequency is maintained at $f = 3\ Hz$. This balance between the two active time scales (precisely $t_u$ and $t_m$) signifies that the MNPs move between the two magnetically active regions in an optimum possible manner. It is worth mentioning here that the optimal movement of MNPs ensures substantial disturbances in the bulk fluid domain, as discussed next. In figure 6, we show the variation of the velocity vectors of the bulk fluid flow inside the ferrofluid droplet domain obtained at different temporal instants of the magnetic actuation cycle with a frequency maintained at $f = 3\ Hz$. As we have already observed from the bright field visualization that, on the actuation of the electromagnets, the motion of the MNPs leads to the development of oppositely directed motion of the bulk carrier fluid. This typical agitation of the bulk flow is the consequential effect of the piercing action of the MNPs on the carrier fluid, as discussed before. Due to this piercing action of the MNPs, a high-pressure zone is created ahead of the MNPs cluster, leading to the development of low pressure behind it. As a consequence of the spatial pressure gradient in the droplet domain, the bulk liquid moves from the high-pressure zone to the low-pressure zone. Following this phenomenon, we observe in figure 6 the bulk flow motion in the opposite direction of the MNPs motion. It is to be mentioned here that $t^* = 0^+$ denotes the state when the magnet is just switched ON, while $t^* = 0.25{:}\,0.5{:}\,0.75$ represents the subsequent intermediate stages. Note that at $t^* = 0^+$, the MNPs realize the magnetic force and start rearranging themselves along the direction of the applied forcing environment. Primarily due to this rearrangement of the magnetic nanoparticles (MNPs), the low intensity of agitation is produced in the droplet flow domain at $t^* = 0^+$. Consequently, we observe low magnitude velocity at this stage, as witnessed in figure 6. At $t^* = 0.5$, a relatively higher velocity is observed in the droplet domain. This is because the head of the moving MNPs on striking the triple contact line of the magnetically active zone generates a tremendous amount of agitation in the droplet fluid domain. Following this impact (of the MNPs with the triple contact line at the magnetically active region), the MNPs rearrange themselves according to the applied magnetic force field. As the MNPs start rearranging themselves, the viscous force of the bulk liquid subsequently dissipates the generated disturbances in the carrier fluid. Primarily due to this reason, we observe a spontaneous drop in the ferrofluid droplet flow velocity at $t^* = 0.75$, as can be seen in figure 6. Similar characteristics can be observed even when the right magnet is turned into an 'On' state, barring the fact that the fluid motion is in the opposite direction to the scenarios pertinent to the left magnet case.

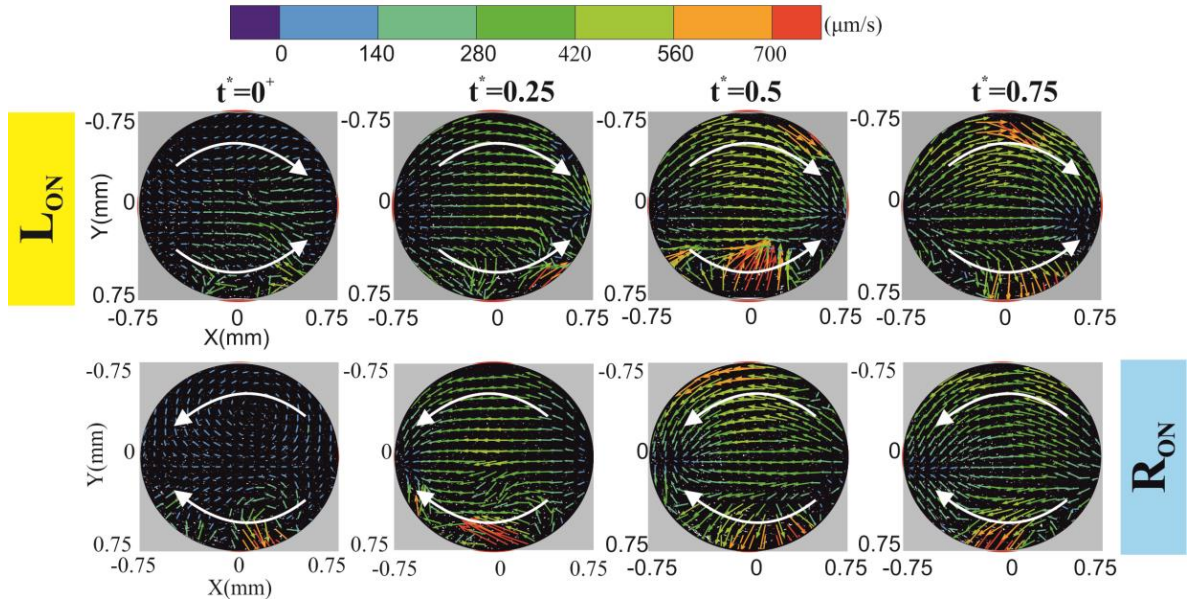

**FIGURE** 6. **(Color online)** Plot depicts the bulk flow motion inside the droplet at various time of operation of the magnetic field, when the frequency of the magnetic field is maintained at 3Hz. The total time of operation of an individual magnet is divided into four parts, with each individual time-steps denoting an increment of $T/4$, where 'T' represents the time of operation of an individual magnet, i.e., $T = 1/2f$. $L_{ON}$ and $R_{ON}$ signifies the state when the left electromagnet and the right electromagnet is active, respectively. The white colored arrows indicate the direction of the bulk liquid flow motion.

To obtain a clearer insight on this distinctive variation in the bulk flow velocity magnitude in the ferrofluid droplet domain, we depict figure 7(a). Note that figure 7(a) shows the variation of $\phi$, which is given as:

$$\phi = \frac{U}{U_{0^+}} \tag{3.1}$$

While $U$ refers to the strength of the velocity inside the droplet domain. $U_{0^+}$ implies the strength of the flow-field at $t^* = 0^+$. The strength of the velocity $(U)$ was calculated by the root mean square of all the velocity vectors along the X-Y plane, and given as:

$$U = \frac{1}{N \times M} \sum_{i=1, j=1}^{N,M} \sqrt{u(i,j)^2 + v(i,j)^2} \qquad (3.2)$$

While $N$ and $M$ refer to the number of grid points in the $X$ and $Y$ directions, respectively. Note that $u(i,j)$ and $v(i,j)$ refers to the instantaneous $X$ and $Y$ directional velocities. From the observation made in the context of bright field investigation, we can recall that the movement of MNPs towards the magnet induces fluid motion inside the droplet. However, the MNPs on reaching further downstream impacts the triple contact line nearer the active magnet, which in turn creates a bulk disturbance in the carrier liquid. This disturbance, however, decays with time because of the effective viscosity of the carrier fluid. Keeping these inferences in mind, we would like to discuss the temporal variation of $\phi$ for the magnetic field frequencies of 0.3 Hz, 1 Hz, 3 Hz, and 5 Hz, respectively, as plotted in figure 7(a). For all the frequencies except for $f = 5\ Hz$, $\phi$ exhibits a positive (+ve) slope, and on reaching its peak value, it encounters a negative (-ve) slope. When the magnetic field frequency is maintained at $f = 5\ Hz$, an almost constant slope is encountered. Also, we report another important observation from figure 7(a) is that, for the $f = 3\ Hz$ case, the $\phi$ curve demonstrated maximum value.

The positive (+ve) slope signifies the rise in the bulk flow velocity, while a negative (–ve) slope of $\phi$ signifies a decrease in bulk flow velocity in the droplet domain. As discussed before, when the electromagnet is actuated, the MNPs moves towards the magnetically active zone. As a consequence of these disturbances, we observe a positive (+ve) slope of $\phi$ during initial temporal instants of the actuation cycle in figure 7(a). Next, when the moving MNPs impacts the triple contact line near the active magnet region, tremendous agitation is produced in the droplet flow field. At this point, we observe a peak velocity in the droplet domain, as witnessed in figure 7(a). Once the velocity reaches a peak value, the viscous effect of the fluid dissipates the agitated flow velocity to the surrounding fluid. Notably, as a consequence of this dissipating effect, following this peak value, we observe a negative (–ve) slope of $\phi$ in figure 7(a).

From the above discussion, it is clear that the viscous force acts as the suppressing agent for the agitation being developed in the flow field. Notably, we observed the presence of critical frequency $(f = 3\ Hz)$ at which the magnetic perturbation time scale almost balances the advective time scale of the MNPs. Following this balance between the dominant time scales, the suppression rate of the agitation intensity is largely reduced. As a result, we observe augmented flow velocity throughout the magnetic cycle for $f = 3\ Hz$ case, which is supported by $\phi > 1$ in figure 7(a). Conversely, for the cases of $f = 0.3\ Hz$ and $f = 1\ Hz$, the magnetic perturbation time scale becomes more significant than the advective time scale. Because of the dominating effect of the perturbation time scale, the disturbances initiated in the domain almost diminishes at the end of the magnetic field cycle. Consequently, we observe in figure 7(a) the value of $\phi$ lower than one ($\phi < 1$) towards the end of the magnetic field cycle. When the magnetic field frequency ($f$) is maintained at 5Hz, the advective time scale ($t_u$) becomes higher than the perturbation time scale. Important to mention here that, because of this difference in involved time scales, the MNPs cannot fully impact both the triple contact line of the droplet at the magnetically active region. This temporal effect leads to a reduction in spatial dispersion of the MNPs in the flow domain. Thus, at this frequency, the motion of the MNPs is highly localized, as seen in figure 5 (see movie 1 given in the supplementary information section). Notably, as a consequence of this phenomenon, we encounter a constant slope for the 5 Hz case in figure 7(a).

In figure 7(b), we show the variation of $\phi_m$ in the droplet domain for a particular cycle of operation of the magnetic field. For the sake of completeness, we here define $\phi_m$ as given by,

$$\phi_m = \frac{\int_t^{t+T} \phi(t) dt}{T} \qquad (3.3)$$

As discussed before, when the magnetic field frequency is maintained at 3 Hz, the advective time scale almost balances the magnetic perturbation time scale. Thus, this frequency ($f = 3Hz$) serves as the critical frequency ($f_{cr}$) at which maximum agitation and minimum dissipation of the fluid velocity takes place in the droplet flow field. Primarily because of this reason, we encounter a high value of $\phi_m$, as can be seen from figure 7(b). Whereas for lower frequencies (particularly, for $f = 0.3\ Hz$ and $f = 1\ Hz$), since the perturbation time scale is very large as compared to the advective time scale ($t_u < t_m$), the disturbances created in the flow domain by the moving MNPs get dissipated well in the field. As a result of this, we observe lower values of $\phi_m$ in figure 7(b). From the ongoing discussion, it may be inferred that, for frequencies higher than critical frequencies, i.e., $t_u > t_m$, the motion of MNPs are highly localized, thereby creating localized agitations in the bulk liquid domain. Due to this limited disturbance, a lower value of $\phi_m$ is encountered at this frequency (cf. figure 7(b)).

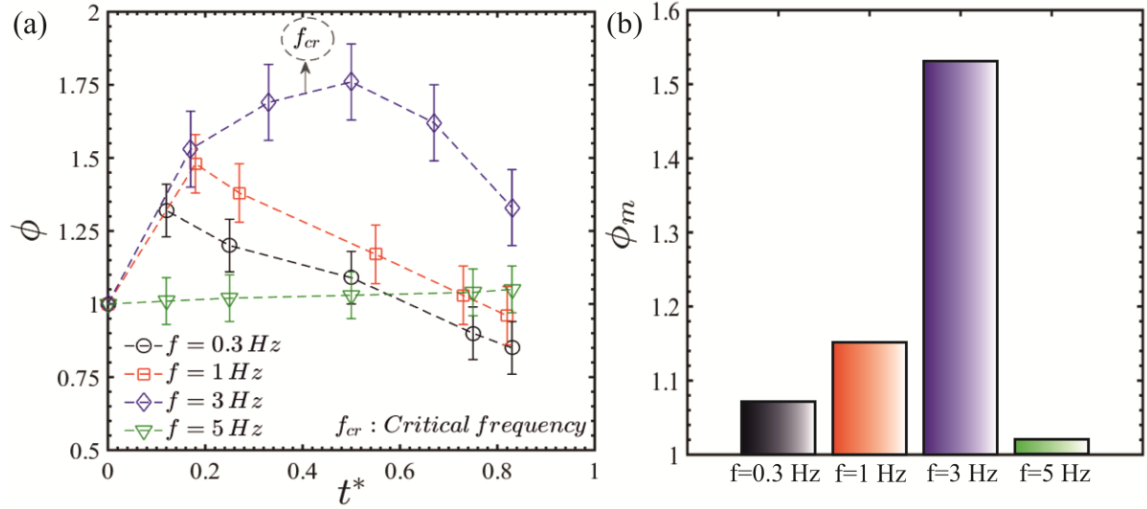

**FIGURE 7. (Color online)** (a) Plot illustrates the temporal variation of $\phi$ for various magnetic field frequencies. The black-colored arrow identifies the critical frequency of the applied magnetic field at which maximum disturbances are produced in the droplet domain. (b) The bar graph depicts the variation of $\phi_m$ for the various magnetic field frequencies.

In order to gain further insights into the underlying flow dynamics, we undertake an effort to predict the circulation produced inside the droplet flow field. In doing so, we calculate the vorticity ($\omega_z$) of the velocity field as,

$$\omega = \frac{1}{A}\Gamma_{ij} \quad (3.4)$$

Where $\Gamma_{ij}$ is the circulation of the flow field, and is given by,

$$\Gamma = \oint (u,v).dl \quad (3.5)$$

In figure 8, we show the variation vorticity contours in the ferrofluid droplet flow field under the influence of various magnetic field frequencies. Two oppositely directed vortices are clearly observed for all the investigating cases. Quite intuitively, the magnitude of vorticity is maximum for the 3 Hz case in comparison to all other cases under consideration. This behavior is in agreement with our previously made observation on the development of maximum augmentation in velocity (cf. figure 8) in the droplet flow field at $f = 3\ Hz$.

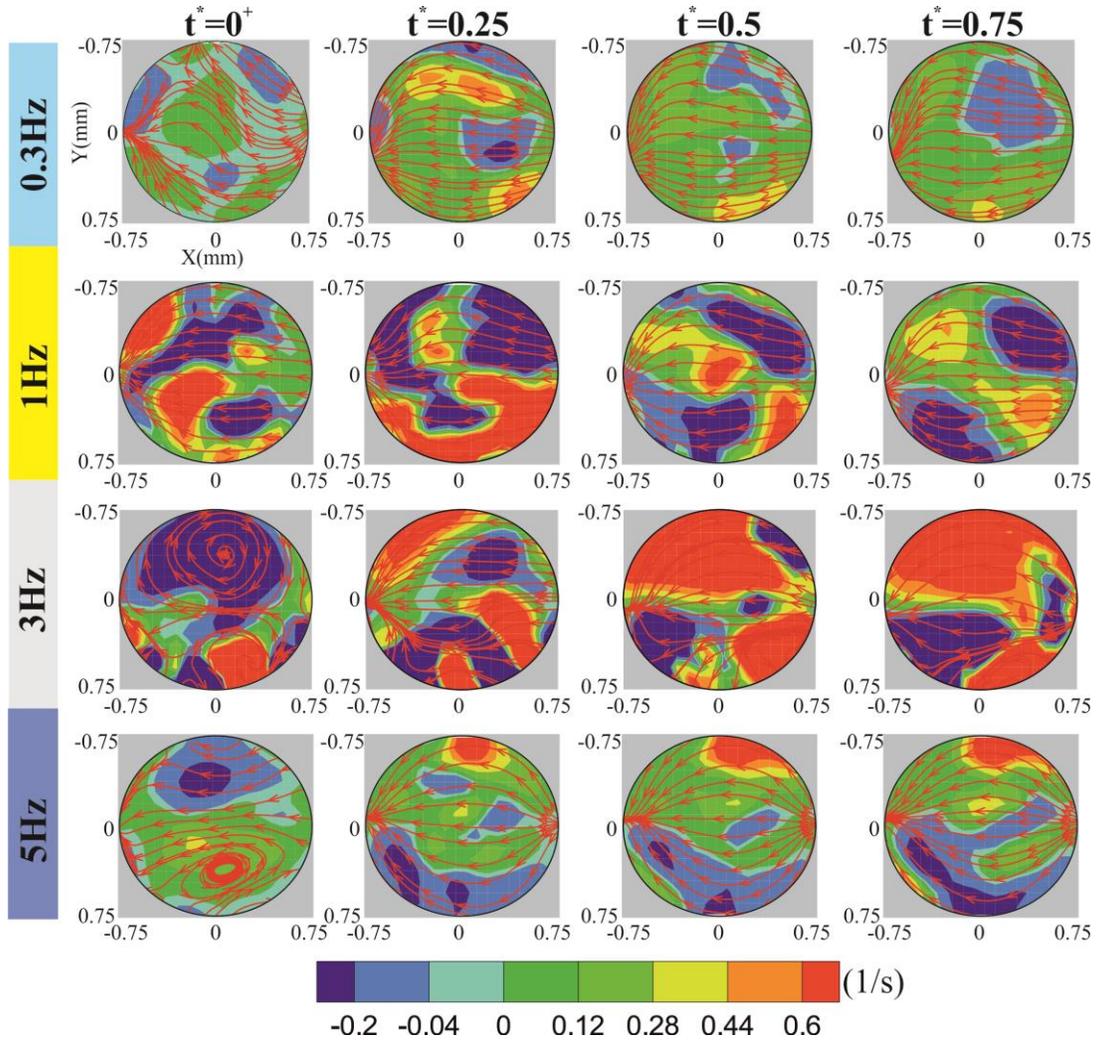

**FIGURE** 8. **(Color online)** Plot depicts the temporal variation of the vorticity contours for the various magnetic field frequency of 0.3 Hz, 1 Hz, 3 Hz, and 5 Hz, respectively. The red-colored arrows indicate the streamlines of the flow. The plot shows the vorticity flow field when the right magnet is in ON state.

### 3.2. Droplet mixing characteristics

*3.2.1. Experimental insights*

From the discussion made in the preceding section, it is apparent that the manipulation of internal convections inside the ferrofluid droplet is possible by careful maneuvering of the MNPs movement. In this section, we explore the role of these convections on the underlying mass transfer between two droplets. The procedure adopted for the mixing process is already described in the materials and methods section.

In the absence of any external force, mixing between two droplets occurs solely due to molecular diffusion. However, the influence of a magnetic field in the paradigm of mixing dynamics at the microfluidic scale leads to a completely different scenario (Hejazian et al. 2016; Zhu & Nguyen 2012). Quite notably, the mixing dynamics gets further amplified in the presence of a time-dependent magnetic field. The amplification is primarily due to the substantial agitation produced in the droplet flow domain under the influence of a time-dependent magnetic field. The time-periodic magnetic actuation leads to interfacial instability, which, in turn, enhances subsequent mixing following an augmented agitation in the droplet flow field. Note that the intensity of this agitation is directly related to the frequency of the applied magnetic field, attributed primarily to its considerable effect on the interfacial instability. Because of this frequency modulated agitation, the underlying mixing between two droplets enhances.

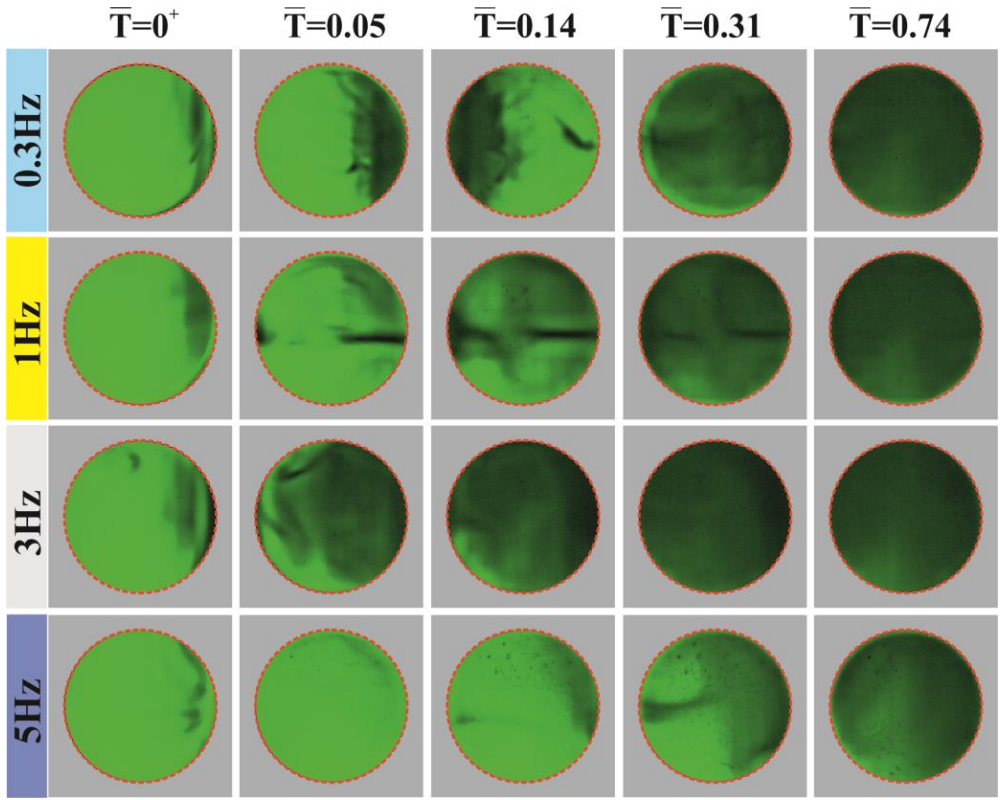

**FIGURE** 9. **(Color online)** Fluorescein distribution inside the droplet flow field at various time instances of the mixing processes for the magnetic field applied frequency of 0.3 Hz, 1 Hz, 3 Hz, and 5 Hz, respectively. $\bar{T}$ represents the non-dimensionalized mixing time and is given as $\bar{T} = t/T_0$ where, $t$ is the instantaneous time and $T_0$ is the total time of mixing between the two droplets in the absence of a magnetic field.

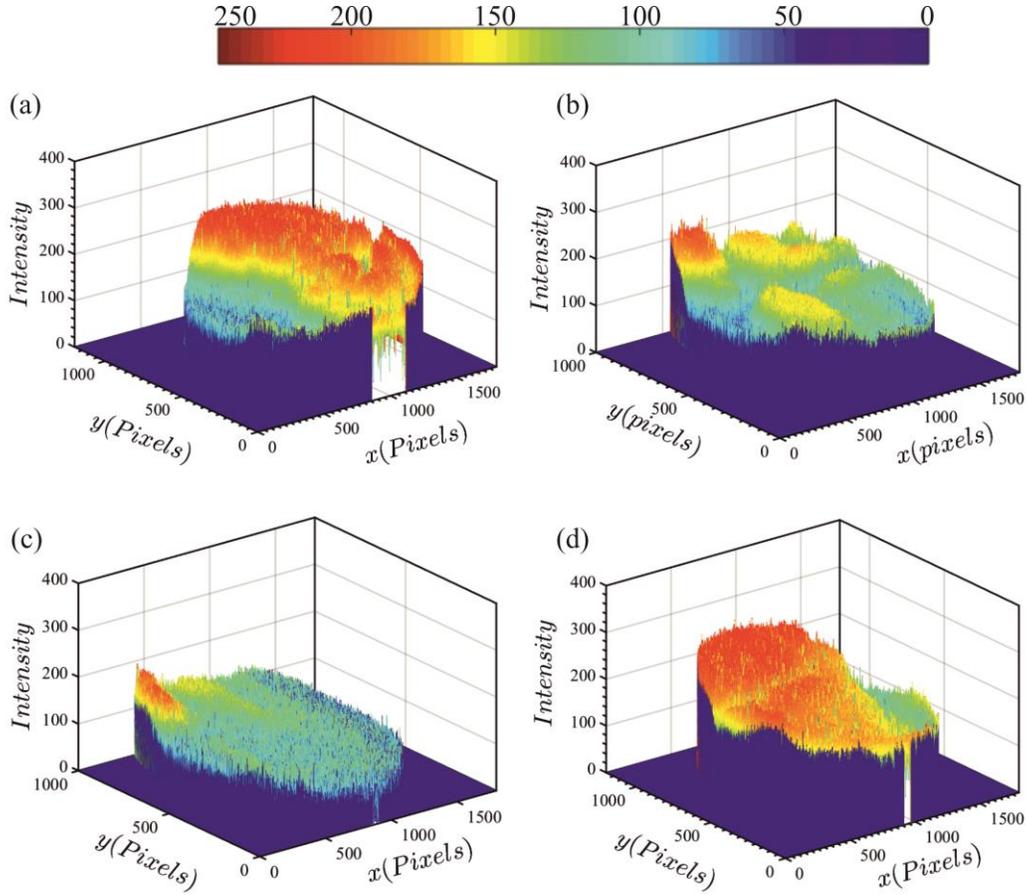

**FIGURE** 10. **(Color online)** Representative distribution of the fluorescence intensity at $\bar{T} = 0.14$ for the magnetic field frequency of (a) $f = 0.3\ Hz$ (b) $f = 1\ Hz$ (c) $f = 3\ Hz$ (d) $f = 5\ Hz$. The color bar shows the pixel intensity ranging from 0 to 255.

In figure 9, we show the distribution of the fluorescein intensity in the flow domain of the mixed droplet at different temporal instants. Important to mention here that a fully mixed state will have a uniform intensity distribution throughout the droplet flow field. Note that $\bar{T}$ represents the non-dimensional mixing time and is defined as $\bar{T} = t/T_0$, where $t$ is the instantaneous time and $T_0$ is the total time of mixing between the two droplets in the absence of the magnetic field. In the presence of a magnetic field, the susceptibility mismatch between the two fluid leads to the development of instability at the liquid-liquid interface. As a result of this instability and its subsequent effect on bulk flow agitation, the overall mixing time between the two droplets is reduced substantially. From figure 9, we can see that the inhomogeneity in the fluorescence distribution gradually reduces over time. A closer observation of figure 9 is suggestive of a better mixing for $f = 3Hz$ case, as realized by a uniform fluorescein field in the domain even at earlier temporal instants $\bar{T} = 0.05$ and $0.14$. Since the severe agitation in the flow field leads to an enhancement in mixing, this observation (fluorescein field for $f = 3\ Hz$) is in support with our argument of augmented flow velocity (which is the effect of severe agitation) for $f = 3Hz$ as discussed before. Thus, it can be argued that the mixing phenomena taking place inside the droplet flow field are a strong function of the frequency of the applied magnetic field (see the movie (2)-(5), given in the supplementary information section). To further ascertain the mixing phenomena occurring between two droplets (precisely, between two fluids) in figure 9, we plot the distribution of the fluorescence intensity at $\bar{T} = 0.14$ in figure 10. We can clearly visualize from figure 10(c) an almost uniform fluorescein distribution inside the droplet flow field when the magnetic field frequency is maintained at $f = 3Hz$. Whereas substantial inhomogeneity exists in the droplet domain for the cases of $f = 0.3\ Hz,\ 1\ Hz,\ and\ 5\ Hz$.

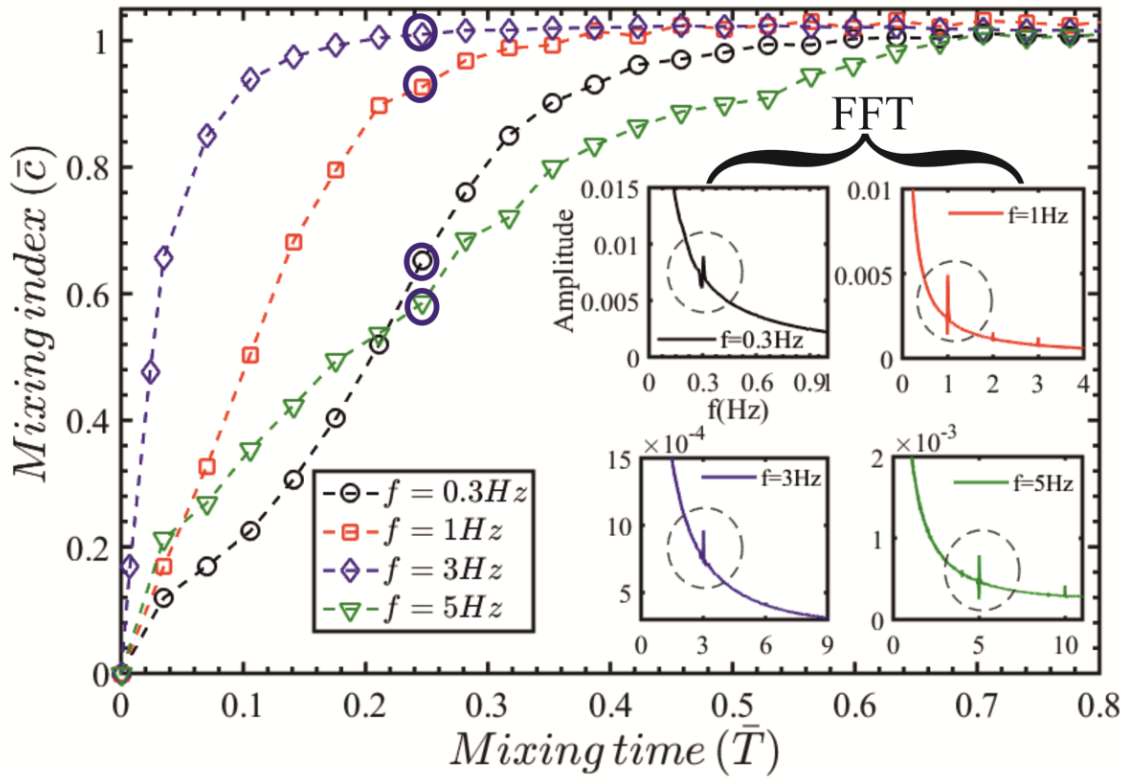

**FIGURE** 11. **(Color online)** Plot depicts the temporal variation of the mixing index ($\bar{C}$) of the droplet flow field for the magnetic field applied frequency of 0.3Hz, 1Hz, 3Hz, and 5Hz, respectively. The insets depict the Fast Fourier Transform of the mixing index for all the cases under consideration. The black color dotted circle highlights the peak of the Fast Fourier Transform curve. The blue color circle identifies the mixing index ($\bar{C}$) at $\bar{T} = 0.25$.

It is established by now that the perturbation frequency ($f$) of the magnetic field plays a dominant role in the magnetofluidic mixing of the microdroplet. In figure 11, we show the variation of the mixing index ($\bar{C}$) versus the non-dimensionalized mixing time, $\bar{T}(= t/T_0)$ for all the cases under consideration. The plot in figure 11 shows that for $\bar{T} = 0.25$, the mixing index ($\bar{C}$) varies as 0.62, 0.94, 0.99, and 0.58 (highlighted by encircled points) for the magnetic field actuation frequencies of 0.3 Hz, 1 Hz, 3 Hz, and 5 Hz, respectively. This particular insight signifies that rapid mixing is possible when the magnetic field frequency ($f$) is maintained at 3 Hz. The inset of figure 11 shows the Fast Fourier Transform (FFT) of the mixing index data. These FFT values justify that the applied magnetic field frequency is the dominant perturbing force acting on the droplet domain.

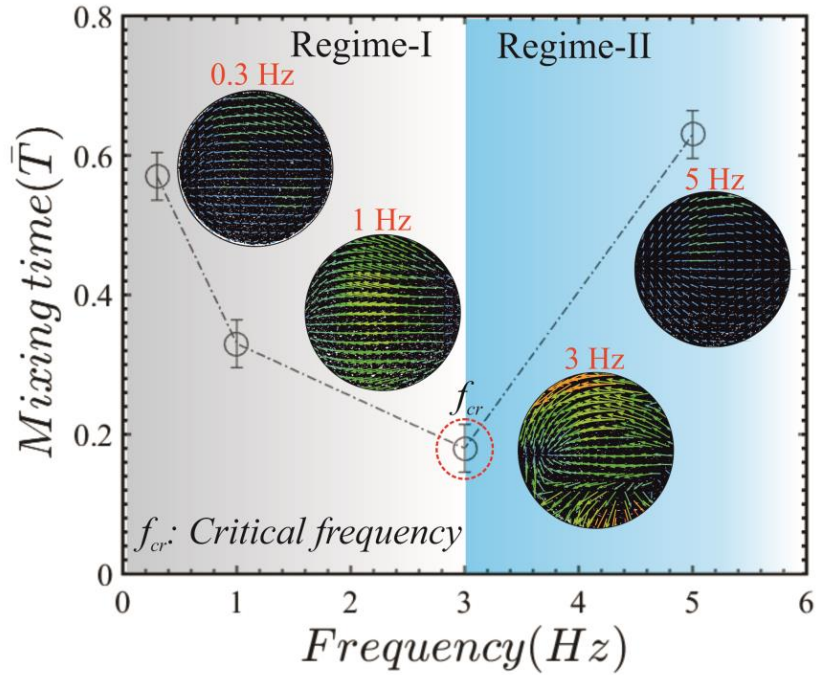

**FIGURE** 12. **(Color online)** Plots depict the experimental variation of the overall mixing time between the two droplets for the various perturbing magnetic field frequency($f$). The black color hollow circle (O) indicates the experimental droplet mixing time at a particular frequency. The critical frequency is identified by the dotted red color circle. Regime-I indicates the zone in which mixing time is inversely related to magnetic field frequency. Regime-II indicates the zone in which the droplet mixing time is directly related to frequency of the magnetic field. The inset shows the velocity distribution inside the ferrofluid droplet domain for all the magnetic field perturbing frequencies.

In the preceding discussion, we have identified the presence of a critical frequency at which the time of the complete mixing process of the two droplets is minimum. Thus, the interactive role of the involved time scales viz., the advective, perturbation, and diffusive time scales on the mixing process is non-trivial as apparent from the ongoing discussion. We have previously seen that the advective and perturbation time scales are almost equal for $f = 3\ Hz$ case. Intuitively, the minimum time for complete mixing in the present scenario should be for $f = 3\ Hz$ case. In figure 12, we show the variation of the overall non-dimensionalized mixing time ($\bar{T}$) between the two droplets for the various frequencies of the perturbing magnetic field. The curve in figure 12 exhibits an initial negative (-ve) slope, and after reaching a critical value, it encounters a positive slope (+ve). As such, the curve in figure 12 can be divided into regime I and regime II, respectively. In regime-I, with an increase in the frequency of the perturbing magnetic field, the overall mixing time between the two droplets reduces. While in regime-II, an increase in frequency of the magnetic field from the critical one, increases the mixing time between the droplets. The critical frequency is encountered at the transition point from regime 1 to regime 2. At the critical frequency ($f_{cr}$), the flow encounters minimum mixing time. This minimum mixing time is primarily due to the aggravated agitations the droplet domain experiences at the critical frequency (cf. inset of figure 12).

*3.2.2. Numerical perspectives*

In the present work, we have used bright field visualization, μPIV, and μLIF to comprehensively explore the droplet mixing characteristics. However, the adopted experimental methodologies have its limitations in observing the instabilities in the concentration flow field of the droplet. In addition to that, the designed circuit of the electromagnet has its restrictions at very high magnetic field frequencies. Primarily because of these reasons, numerical simulations are conducted in COMSOL Multiphysics® to explore the concentration field of the mixed droplet at higher frequencies. Also, we took advantage of simulations to have a qualitative understanding of the instabilities occurring in the droplet domain. In figure 13, we compare the simulated results of the mixing index ($\bar{C}$) with our experimental data in absence of any external magnetic forcing. A good match between the experimentally observed and numerically calculated mixing index justifies the reliability of our experiments. The inset of figure 13 shows the instantaneous numerical and experimental evolution of the droplet concentration field. The present numerical modeling framework, albeit 2D, is benchmarked with the experimental data and can be used as a tool to extract detailed insight in the droplet mixing physics.

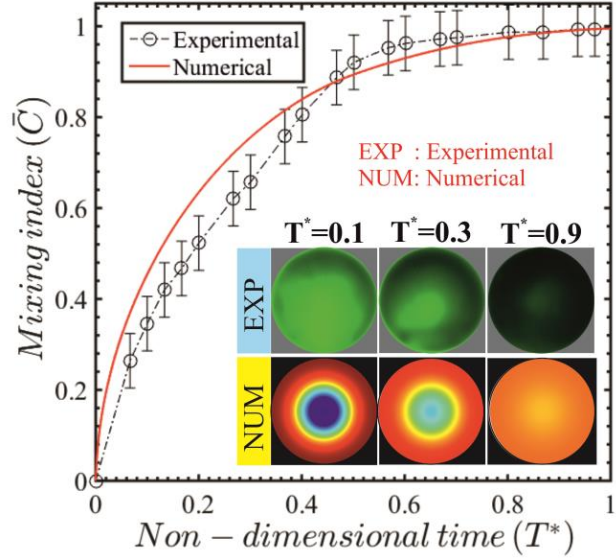

**FIGURE** 13. **(Color online)** Plot benchmarks the experimentally calculated mixing index with that of the numerically calculated mixing index, in the absence of an external magnetic field. $T^* = t/T_0$, where $T_0$ is the total time of mixing of the droplet. The inset shows the snapshots of the temporal evolution of the concentration flow field of the droplet.

We have discussed about the interfacial instability of two fluids (miscible) and argued its effect on the underlying mixing in the preceding section. It is worth mentioning here that the instability is induced in the droplet flow domain due to the magnetic susceptibility mismatch between the two fluids. These disparities in the effective magnetization ensure an increase in the overall interfacial area leading to the development of "*finger-like*" fronts, as can be observed from figure 14(a). Note that the "*finger-like*" fronts appearing at the interface are an indicative measure of the interfacial instability. We show in figure 14(a) the temporal evolution of the concentration field for various perturbing frequencies. Although the prime aim of the present endeavor is not to critically analyze the instability picture, yet we have discussed the "*finger-like*" fronts being developed at the interface during magnetic perturbation for a broader understanding of the mixing dynamics in the present scenario. It is worth mentioning here that the instability picture, as discussed above, is not the numerical artifact; instead, the appearance of the *"finger-like"* structures is the consequence of the mismatch of the magnetic susceptibility of two fluids. Readers are referred to the supplementary material part (Section 3) of this paper, wherein the detailed analysis of the concentration field with a finer grid size is presented to support this conjecture. With the alteration in the frequency of the magnetic field, the distinctive spatial variation of the concentration flow domain is apparent. A closer inspection of figure 14(a) reveals that the numerical results showing uniform distribution of the concentration field for $f = 3Hz$ are in coherence with the experimental observations, as demonstrated in figure 9. Moreover, at higher magnetic field frequencies, the droplet domain attains an almost constant behavior. This constant behavior is primarily due to the fact that at a higher perturbing frequency, the magnetic perturbation time scale is very low as compared to the advective time scale of the flow, i.e., $t_m \ll t_U$. Consequently, under the application of very high magnetic field frequencies, any perturbation could not be fully propagated, and the droplet domain behaves as if it is acted upon by two magnets under steady operations. Figure 14(b) shows the numerical and experimental evolution of the concentration flow field at the critical frequency of 3Hz. Consistency in the concentration flow field between the experimental and numerical results, as observed in figure 14(b), justifying the reliability of the modeling framework developed in this analysis.

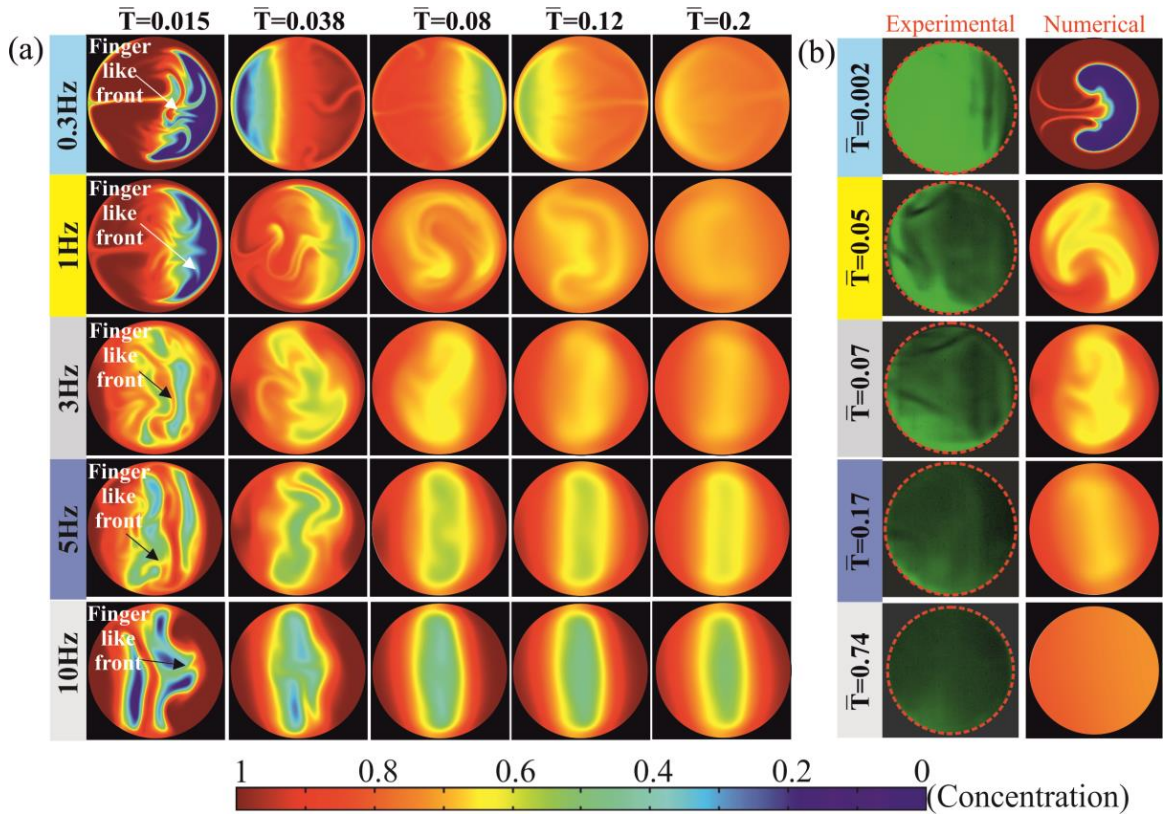

**FIGURE 14. (Color online)** (a) Concentration distribution inside the droplet flow field at various temporal instances of the droplet mixing process for the magnetic field applied frequency of 0.3Hz, 1Hz, 3Hz, 5Hz, and 10Hz respectively. The black and white-colored arrow identify the finger-like front developed in the droplet domain due to the magnetization differences between the two fluids. (b) Plots depict the temporal evolution of the experimental and numerical variation of the concentration flow field when the magnetic field frequency is maintained at 3Hz.

4. **Concluding remarks**

In summary, we report the experimental investigations of the mixing dynamics of a ferrofluid droplet with a non-magnetic droplet under the influence of a time-dependent magnetic field. We show that the intermittent motion of magnetic nanoparticles (MNPs) under the external forcing induces a magneto convective flow inside the ferrofluid droplet. By performing the bright field visualizations, we obtain the qualitative understanding of the MNP's motion inside the ferrofluid droplet, while μPIV investigation is carried out for quantification of the bulk flow dynamics inside the domain. As observed, the flow convection inside the ferrofluid droplet gets augmented in the presence of a time-dependent magnetic field. We numerically simulate the flow dynamics inside the ferrofluid droplet domain and explore the existence of interfacial instability, which initiates the mixing in the present problem. A mismatch of the magnetic susceptibility of two fluids, together with the viscosity contrast, triggers the mixing in the convective mixing regime. A critical frequency is observed at which the internal convection inside the droplet is amplified in the presence of a magnetic field. At this critical frequency, the advective time scale of the flow is balanced by the magnetic perturbation time scale. This balance ensures optimal reciprocation of the MNPs in between the two magnetically active zones. At a lower frequency, the residence time of the MNPs at a particular magnetically active zone increases, ensuring that the agitated energy of the bulk flow is dissipated by the viscous energy of the flow. While at a higher frequency, the MNPs are unable to reach the magnetically active zone, thereby restricting the agitation developed in the bulk flow to a particular limit. Since the agitation developed in the droplet domain is maximum at the critical frequency, the time of complete mixing between the two droplets becomes minimum at this frequency. We show that the critical frequency obtained from the experimental observations is in good agreement with the numerical result. At the critical frequency, the overall mixing time between the two droplets is reduced by almost 80% when compared with the base case, i.e., no applied magnetic field. We believe the proposed technique will enable numerous biomicrofluidic and Lab-on-a-CD based applications towards achieving efficient mixing in a less invasive way.

**Acknowledgment**

SS and PKM acknowledge the experimental facilities of Microfluidics and Microscale Transport Processes Laboratory in the Mechanical Engineering Department at IIT Guwahati. The authors acknowledge the CIF, IIT Guwahati, for the support in the characterization of ferrofluid.

**Conflict of interest**

The authors declare no conflict of interest.